\documentclass[aps,pre,groupedaddress,twocolumn]{revtex4}
\pdfoutput=1

\usepackage{hyperref}
\usepackage{subfigure}
\usepackage{color}
\usepackage{graphicx}
\usepackage{natbib}
\usepackage{doi}

\begin{document}
\bibliographystyle{unsrt}

\expandafter\ifx\csname urlprefix\endcsname\relax\def\urlprefix{URL }\fi
\DeclareGraphicsExtensions{.pdf, .jpg}

\title{\large
Dark matter is a manifestation of the vacuum Bose-Einstein condensate
}

\large
\author{Valeriy I. Sbitnev}
\email{valery.sbitnev@gmail.com}
\address{St. Petersburg B. P. Konstantinov Nuclear Physics Institute, NRC Kurchatov Institute, Gatchina, Leningrad district, 188300, Russia;\\
 Department of Electrical Engineering and Computer Sciences, University of California, Berkeley, Berkeley, CA 94720, USA
}

\date{\today}

\begin{abstract}
The vorticity equation stemming from the modified Navier-Stokes equation gives a solution for a flat profile of the orbital speed of spiral galaxies.
Solutions disclose existence of the Gaussian vortex clouds, the coherent vortices with infinite life-time, what can be a manifestation of the dark matter.
The solutions also disclose what we might call a breathing of the galaxies - due to an exchange of the vortex energy with zero-point fluctuations in the vacuum.



\end{abstract}

\maketitle

\large

\section{\label{sec=1}Introduction}

\begin{figure}[htb!]
 \centering
  \begin{picture}(200,140)(0,10) 
      \includegraphics[scale=0.5, angle=0]{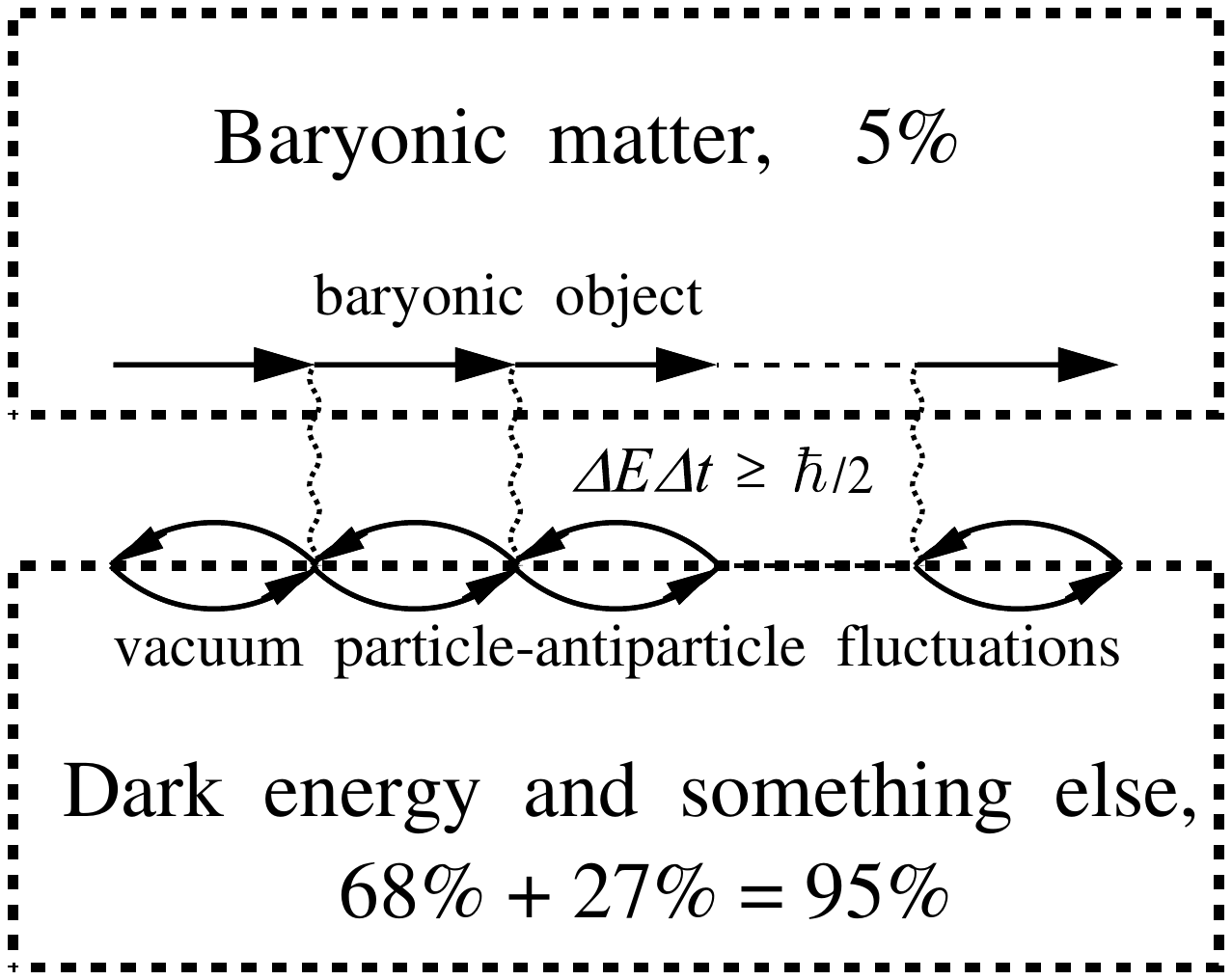}
  \end{picture}
  \caption{
Baryon particles move over a vast ocean of the dark energy and something else named the dark matter.   }
  \label{fig=1}
\end{figure}

The belief in the existence of an aether~\citep{Whittaker1990} - a world medium filling all of space, which is the building material for all kinds of substances and movements manifesting themselves as force fields - has accompanied natural science from the most ancient times up to now. The building material we currently admit is ordinary (baryonic) matter, but this material accounts only for~5 percent of all matter/energy in the observed Universe. Factually, this matter moves through a vast ocean of the dark energy ($\sim$68 \%) and something else, Fig.~\ref{fig=1}. 
 The scientific community believes that unaccounted (invisible) substances is the dark matter ($\sim$27 \%). 
 
 This vast ocean of the dark essence ($\sim$95~percent) is at perpetual movement. Marco Fedi\cite{Fedi2016} proposes to name this ocean as "the superfluid quantum space", metric of which is "flat". It means, it corresponds to the Euclidean geometry~\cite{SorliEtAl2016}. This space contains enormous amount of virtual particle-antiparticle pairs, which arise and annihilate again and again. In effect, it is the Bose-Einstein mega-condensate~\citep{DwornikEtAl2014a, DwornikEtAl2014b, DasBhaduri2015}.

We can account for dark matter in terms of a more fundamental building material if we treat dark matter as a dark fluid~\citep{Ardey2006, ChefranovNovikov2010, Fedi2016}, whose motion mimics a Madelung quantum fluid, governed by the hydrodynamic Euler equations~\citep{BoehmerHarko2007}.  The motion of that fluid could be described by the Gross-Pitaevskii equation~\citep{BoehmerHarko2007, HarkoMocanu2012}.  

From this perspective, the abundance of dark matter is due to the vacuum energy density which fills a non-trivial space-time structure~\citep{AlbaretiEtAl2014a}.
It originates from the quantum vacuum, which is everywhere densely filled by the virtual particle-antiparticle pairs. With this assumption on board, dark matter can be explained by gravitational polarization of the quantum vacuum~\citep{Hajdukovic2011a}. 
The dark matter was originally invoked to correct observation with theory when it comes to the flat profiles we observe in the orbital speeds of spiral galaxies~\citep{Rubin2004}. Observations show that the orbital speeds of the spiral arms of galaxies stay almost constant as distance increases from the galactic cores ~\citep{deBlokEtAl2001}. 

We adopt here the following facts: (i)~based on observations of the Planck Observatory~\cite{PlanckCo2015} it is determined that the parameter of curvature very close to zero. In other words, the universe is flat with high precision $\Omega_{K} = 0.0008\pm0.004$; (ii)~the speed of rotation of the external regions of galaxies is no more than 300 km/s; (iii)~the speeds of galactic and extragalactic particles can reach 10$^3$ km/s and maybe up to $2\cdot10^3$ km/s, at least~\citep{Parkhomov1998}. In other words, the speeds are non-relativistic.
In accordance with these observations, we may apply further the non-relativistic equations in the Euclidean geometry.

The aim of this article is to explain the flat profile of  orbital speeds in spiral galaxies as a natural consequence of a superfluid vacuum ~\citep{Sbitnev2015c}. Sec.~\ref{sec=2} employs the modified Navier-Stokes equation to describe the motion of a special superfluid medium - the physical vacuum~\citep{Sbitnev2015b, Sbitnev2016b}. The vorticity equation stemming from this modified Navier-Stokes equation~\citep{Sbitnev2016c} gives a solution describing the flat profile of orbital speeds. This is followed by concluding remarks and numerical estimations Sec.~\ref{sec=3}. 
 
\section{\label{sec=2}Modified Navier-Stokes equation and motion of superfluid BEC}
 
The modified Navier-Stokes equation is~\citep{Sbitnev2016b}
\begin{equation}
 m\biggl(
 {{\partial {\vec {\mathit v}}}\over{\partial\,t}}
 + ({\vec {\mathit v}}\cdot\nabla){\vec {\mathit v}}
       \biggr) 
  =  {{{\vec{\mathit F}}}\over{N}}
   \;-\; \nabla Q
 \; +\; m\nu(t)\,\nabla^{\,2}{\vec {\mathit v}}.
\label{eq=1}
\end{equation}
This equation describes a velocity field under the influence of different forces acting on this special deformable medium. The difference between this equation and the familiar Navier-Stokes equation is in two last terms, which describe internal forces arising in the medium. The first force,  $\nabla Q = \nabla(P/\rho)$, is the gradient from the internal pressure $P$ divided by the density distribution of the matter,  $\rho$.
The quantum potential $Q$ is a fundamental element that creates a geometrodynamic picture of the quantum world both in the non-relativistic domain~\citep{Sbitnev2015b, Sbitnev2016b}, and in the relativistic domain~\citep{Sbitnev2015c, LicataFiscaletti2014b, Fiscaletti2012, AliDas2015,  DasBhaduri2015}.  
 It is proportional to the internal pressure $P=P_1+P_2$  divided by the density distribution $\rho$ of the matter in the space. The pressures $P_1$ and $P_2$ follows from the Fick's law, ${\vec J} =-mD\nabla\rho$, applied to distribution of the particle-antiparticle condensate inhabiting the vacuum~\citep{Sbitnev2015c,  Sbitnev2016b}:
\begin{equation}  
 \left\{
   \matrix{
      P_1 = D\nabla{\vec J} = -{\displaystyle{{\hbar^2}\over{4m}}\nabla^2\rho},~~~~ \cr
      P_2 
     {\displaystyle ={{\rho}\over{2m}}\Biggr( {{\vec J}\over{\rho}} \Biggl)^2 = {{\hbar^2}\over{8m}}{{(\nabla \rho)^2}\over{\rho}}}. 
             }
   \right.
\label{eq=2}   
\end{equation}
 Here $D=\hbar/2m$ is the diffusion coefficient~\citep{Nelson1966}, and $\hbar$ and $m$
 are the reduced Planck constant and the particle mass, respectively.
 
  So, the quantum potential reads
\begin{equation}
   Q = {{P}\over{\rho}}
   = -{{\hbar^2}\over{4m}}{{\nabla^2\rho}\over{\rho}} + {{\hbar^2}\over{8m}}{{(\nabla\rho)^2}\over{\rho^2}}.
\label{eq=3}
\end{equation}
  The quantum potential looks as the ratio of the pressure to the density distribution of the matter 
  that is proportional to the energy density.
As a consequence, the internal pressure opposes the gravitational attraction. This means that the osmotic pressure pushes the baryonic matter apart. This may be a cause of the formation of voids in the distribution of galaxies~\citep{ClampittJainm2015a}.
    
  The presence of the quantum potential in the Navier-Stokes equation allows us to describe motion of this medium through the complex-valued wave function $\Psi=\sqrt{\rho}\exp\{{\bf i}S/\hbar\}$ to a solution of the non-linear Schr{\"o}dinger equation, or more specifically, a Gross-Pitaevskii-like equation~\citep{Sbitnev2016b}
\begin{widetext}
\begin{equation}
\label{eq=4}
\hspace{-8pt}
  {\bf i}\hbar\,{{\partial\Psi}\over{\partial\,t}}=
  {{1}\over{2m}}(-{\bf i}\hbar\nabla + m{\vec{\mathit v}}_{_{R}})^2\Psi
      + U({\vec r})\Psi  
      + \underbrace{m\nu(t) {{d \ln(\rho)}\over{d\,t}} }_{(a)}\Psi 
       - C\Psi.
\end{equation}
\end{widetext}
Here we write ${\vec{\mathit v}}={\vec{\mathit v}}_{_{S}}+{\vec{\mathit v}}_{_{R}}$, where ${\vec{\mathit v}}_{_{S}}=\nabla S/m$ is the irrotational velocity ($S$ is the action), and ${\vec{\mathit v}}_{_{R}}$ is the rotational (solenoidal) velocity.
 The term  $m{\vec{\mathit v}}_{_{R}}$ represents the angular momentum of a swirling motion.
 An extra term enclosed by the curly bracket (a) fluctuates about zero due to the fluctuating about zero parameter $\nu(t)$. 
  The potential energy $U({\vec r})$  comes from the expession $\nabla U = -{\vec F}/N$, and $C$ is the integration constant.

\subsection{\label{subsec=2A}Vorticity equation and solutions for orbital speeds of spiral galaxies}

 The modified Navier-Stokes equation describes also a motion of the vortex against background of the superfluid BEC medium.
\begin{figure}[htb!]
  \centering
  \begin{picture}(200,160)(0,10)
      \includegraphics[scale=0.35]{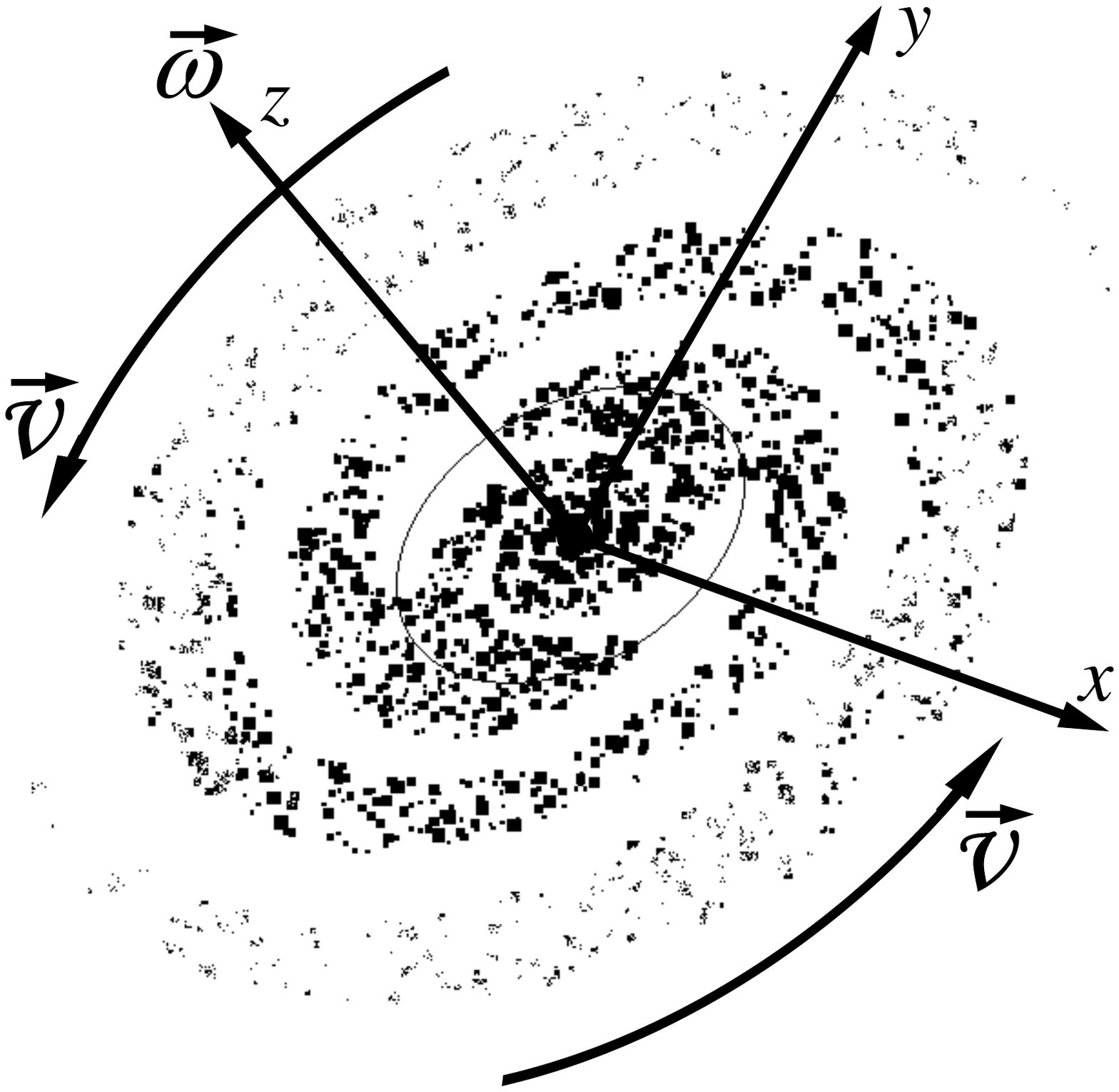}
  \end{picture}
  \caption{
A simulation of a rotating spiral galaxy: the orbital velocity ${\vec{\mathit v}}$ lies in the plane $(x, y)$.
The vorticity vector ${\vec\omega}$ is oriented perpendicular to this plane.    
Orientations of these vectors  ${\vec\omega}$ and ${\vec{\mathit v}}$ obey the right hand rule.
  }
  \label{fig=2}
\end{figure}
   The second internal force in Eq.~(\ref{eq=1}) is dissipative due to a fluctuating presence of a viscosity of the medium. If the viscosity coefficient $\nu$ is a function of time, we can assume that (a) time-averaged, the viscosity coefficient vanishes and (b) its variance is not zero. So the viscosity coefficient is a function fluctuating about zero.
 We suppose that such fluctuations represent an exchange of energy between the existing baryonic matter and zero-point fluctuations of this special superfluid medium - the physical vacuum,  Fig.~\ref{fig=1}.
We can imagine this special medium as the absolute black body at ultra-low temperature with which the baryonic matter is in thermodynamic equilibrium~\citep{HarkoMocanu2012, Poluyan2015R}.

Let us apply the curl operator to the Navier-Stokes equation.
We come to the equation for the vorticity 
${\vec\omega}= [\nabla\times{\vec{\mathit v}}]$~\citep{ProvenzaleEtAl2008, KleinebergFriedrich2013}
\begin{equation}
 {{\partial\, {\vec\omega}}\over{\partial\,t}}
 + (\vec{\mathit v}\cdot\nabla){\vec\omega}
 = \nu(\,t\,)\nabla^{2}{\vec\omega}.
\label{eq=5}
\end{equation}
   This vector is directed along the rotation axis.
  In order to simplify this task, let us move to the coordinate system where the rotation occurs in the plane $(x,y)$ and the $z$-axis lies along the rotation axis~Fig.~\ref{fig=2}.  
  Under this transformation the equation for the vorticity takes a particularly simple form:
 \begin{equation}
  {{\partial\, {\omega}}\over{\partial\,t}} =
 \nu(\,t\,)\Biggl(
    {{\partial^{\,2}\omega}\over{\partial\,r^{2}}}
 +{{1}\over{r}}{{\partial\,\omega}\over{\partial\,r}}
               \Biggr).
\label{eq=6}
\end{equation}
 A general solution of this equation has the following view~\citep{Sbitnev2015c}
\begin{widetext}
\begin{equation}
\label{eq=7}
 \omega(r,t)={{\mit\Gamma}\over{4\Sigma(\nu,t,\sigma)}}
  \exp
  \matrix{
  \left\{\displaystyle
      -{{r^2}\over{4\Sigma(\nu,t,\sigma)}}  
      \right\}},
\end{equation}
\begin{equation}
 {\mathit {v}}(r,t)
= {{1}\over{r}}\int\limits_{0}^{r}\omega(r',t)r'dr' 
={{\mit\Gamma}\over{2 r}}
 \matrix{
             \left( 1 -
  \exp
  \left\{\displaystyle
      -{{r^2}\over{4\Sigma(\nu,t,\sigma)}}  
      \right\}
           \right)
          }.
\label{eq=8}
\end{equation}
\end{widetext}
 The first function is the vorticity and the second is the orbital speed.
 We do not mark the arrows above the letters $\omega$ and $\mathit {v}$, since the vorticity lies on $z$-axis
  and the orbital velocity lies in the $(x,y)$ plane.
    The denominator ${\mit\Sigma}(\nu,t,\sigma)$ in these formulas has the following view
\begin{equation}
   {\mit\Sigma}(\nu,t,\sigma) =
   \int\limits_{0}^{t} \nu(\tau) d\tau + \sigma^{2}.
\label{eq=9}
\end{equation} 
 Here $\sigma$ is an arbitrary constant such that the denominator is always positive. 
 
The extra parameter $\sigma$ ensures the existence of  the Gaussian coherent vortex cloud~\cite{KevlahanFarge1997} having permanent the vorticity  and the angular speed
\begin{eqnarray}
\label{eq=10}
  && \omega_{c}(r) ={{\mit\Gamma}\over{4\sigma^2}}
   \exp\Biggr\{
              -{{r^2}\over{4\sigma^2}}
          \Biggl\}, \\
  && {\mathit v}_{c}(r) = {{\mit\Gamma}\over{2r}}
   \Biggr(
      1 - \exp\Biggr\{
                          -{{r^2}\over{4\sigma^2}}
                  \Biggl\}
   \Biggl)
\label{eq=11}
\end{eqnarray}
 with the circulation $\mit\Gamma$ and the average radius $\sigma$ initially existing.
 Here  the subscript  {\it c}   comes from  the  Gaussian {\it coherent} vortex {cloud}.
Factually, the coherent vortex  represnts localized concentration of energy and vorticity with a life-time tending to infinity~\citep{ProvenzaleEtAl2008}. It does not interact with any form of matter and exists in itself infinitely in time. 
We can express a hypothesis that these Gaussian coherent vortex clouds having a long-time memory
can be a manifestation of the dark matter.  
  
\begin{figure}[htb!]
  \centering
  \begin{picture}(200,80)(-20,10)
      \includegraphics[scale=0.6]{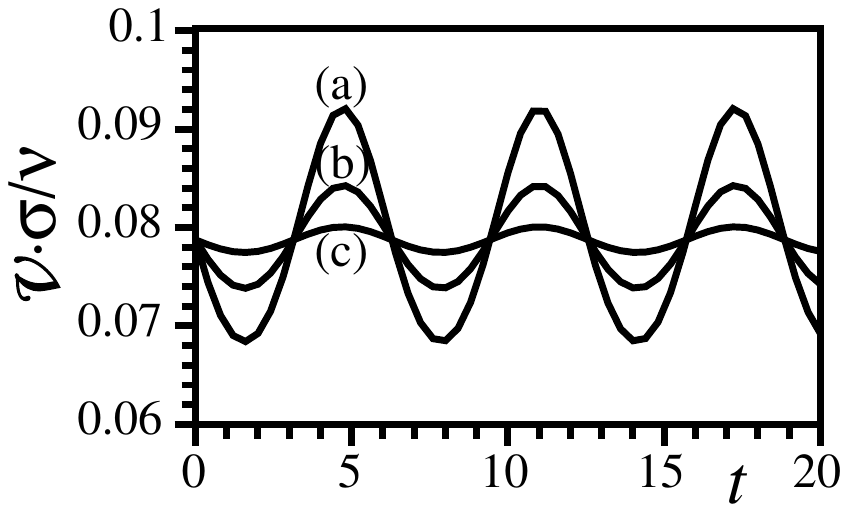}
  \end{picture}
  \caption{Family of oscillations of the orbital speed ${\mathit v}(r,t)$ 
  for different values of $\sigma$:  (a)~$\sigma=2$; (b)~$\sigma=3$; (c)~$\sigma=6$.
  The  factor $\sigma/\nu$  reduces the speed to dimensionless representation.
  }
  \label{fig=3}
\end{figure}

  For the sake of demonstration let us set
\begin{equation} 
 \nu(t) = \nu {{e^{{\bf i}\Omega t} + e^{-{\bf i}\Omega t}}\over{2}} = \nu\cos(\Omega t).
\label{eq=12}
\end{equation}
 In this case
\begin{equation}
 \Sigma(\nu,t,\sigma) = (\nu/\Omega)\sin(\Omega t) +  \sigma^{2}.
\label{eq=13}
\end{equation} 
Fig.~\ref{fig=3} shows family of oscillations of the orbital speed ${\mathit v}(r,t)$ for different values of the parameter~$\sigma$. 
 Variations of the speeds are shown in a region of their maximal changes, at $r = 2\sigma$. 
For bringing the speeds to an equal level, that are calculated at different values of $\sigma$, they are multiplied by the factor $\sigma/\nu$, leading the speeds to dimensionless representation.
 One can see that at increasing $\sigma$ oscillations of the speed diminish. So, at very large $\sigma$ the oscillations vanish.

\subsection{\label{subsec=2B}Hierarchy of the parameters $\nu$ and $\sigma$  and the problem of the flat orbital speed}

\begin{figure}[htb!]
  \centering
  \begin{picture}(200,160)(-5,10)
      \includegraphics[angle=0, scale=0.6]{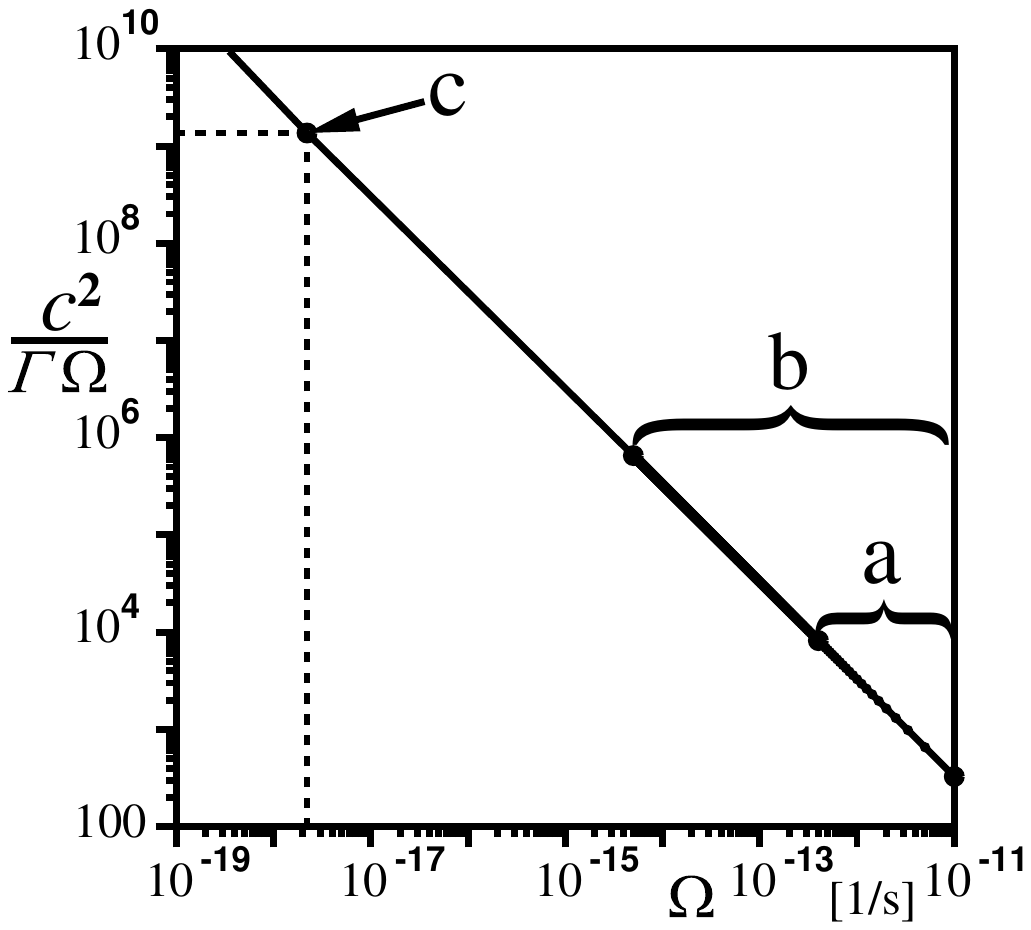}
  \end{picture}
  \caption{
The dimensionless coefficient $c^2/{\mit\Gamma}\Omega$ vs. $\Omega$: (a) $\Omega$ ranges from
$10^{-11}$  s$^{-1}$ to $4\cdot10^{-13}$ s$^{-1}$ for the distance $r$ from the galactic core changing form 0 to 20 kpc, ${\mit\Gamma} \approx 3.2\cdot10^{25}$ m$^2$/s; (b) $\Omega$ ranges from  $10^{-11}$  s$^{-1}$ to $5\cdot10^{-15}$ s$^{-1}$ for the distance $r$ changing form 0 to 40 kpc, ${\mit\Gamma} \approx 2.75\cdot10^{25}$ m$^2$/s.
The arrow points to the limiting edge of the visible universe.
    }
\label{fig=4}
\end{figure}

Observe first that the Gaussian vortex clouds, which manifest themselves like an unknown dark substance, possesses the self-similarity, what can be seen from the above evaluations.
Let us assume in this key that the fluctuating viscosity has the following presentation~\citep{Sbitnev2015c}
\begin{equation}
  \nu_{n}(t) = {{c^{2}}\over{\Omega_{n}}}\cos\{\Omega_{n}t\}.
\label{eq=14}
\end{equation}
 The kinetic viscosity coefficient $c^{2}/\Omega_{n}$ has dimension [length$^2$/time].
 Here $c$ is the speed of light and  $\Omega_{n}$ is
 the angular frequency of a vacuum oscillation.The viscosity obeys to the $1/f$-law (the flicker-noise law) 
 when $\Omega_{n} \sim n^{-1}$ tends to zero at $n$ going to infinity.
From the above formula it follows that the viscosity goes to infinity as $\Omega_{n}$ tends to zero, see Fig.~\ref{fig=4}.  
 In this figure 
  intervals of changing $\Omega_{n}$ are marked by the curly brackets: (a) $\Omega_{n}$ ranges from $10^{-11}$ to $4\cdot10^{-13}$~1/s when $n$ goes from 1 to 25. In this case the flat profile of the orbital speed is observed from about 10 to 20 kpc which has been shown in~\citep{Sbitnev2015c}; (b) $\Omega_{n}$ ranges from $10^{-11}$ to $5\cdot10^{-15}$~1/s when $n$ goes from 1 to 2000. The flat profile observed in this case is from about 10 to 40 kpc.
  So, the lower the frequency $\Omega_{n}$ is taken, the larger radius of the flat profile of the orbital speeds can be achieved. 
 
       De Broglie wavelength, $\lambda_{n}=c/\Omega_{n}$,  in these cases ranges from about 20 kpc to 2 Mpc and covers the galactic scales. 
We can evaluate masses, $m = {{\hbar\Omega_{n}}/{c^{2}}}$, for the axion-like particles~\citep{MielczarektAl2010} which range from about $10^{-62}$~kg  to $5\cdot10^{-66}$~kg. They are in the range shown in~\citep{MureikaMann2011}.

\subsection{\label{subsec=2B}Solution for a flat profile of the orbital speed}

 A rich gallery of the galaxy rotation curves showing output on a flat profile is presented in~\citep{deBlokEtAl2001}.
 Let us try to compute the flat profile.

\begin{figure}[htb!]
  \centering
  \begin{picture}(210,220)(20,10)
      \includegraphics[angle=0, scale=0.45]{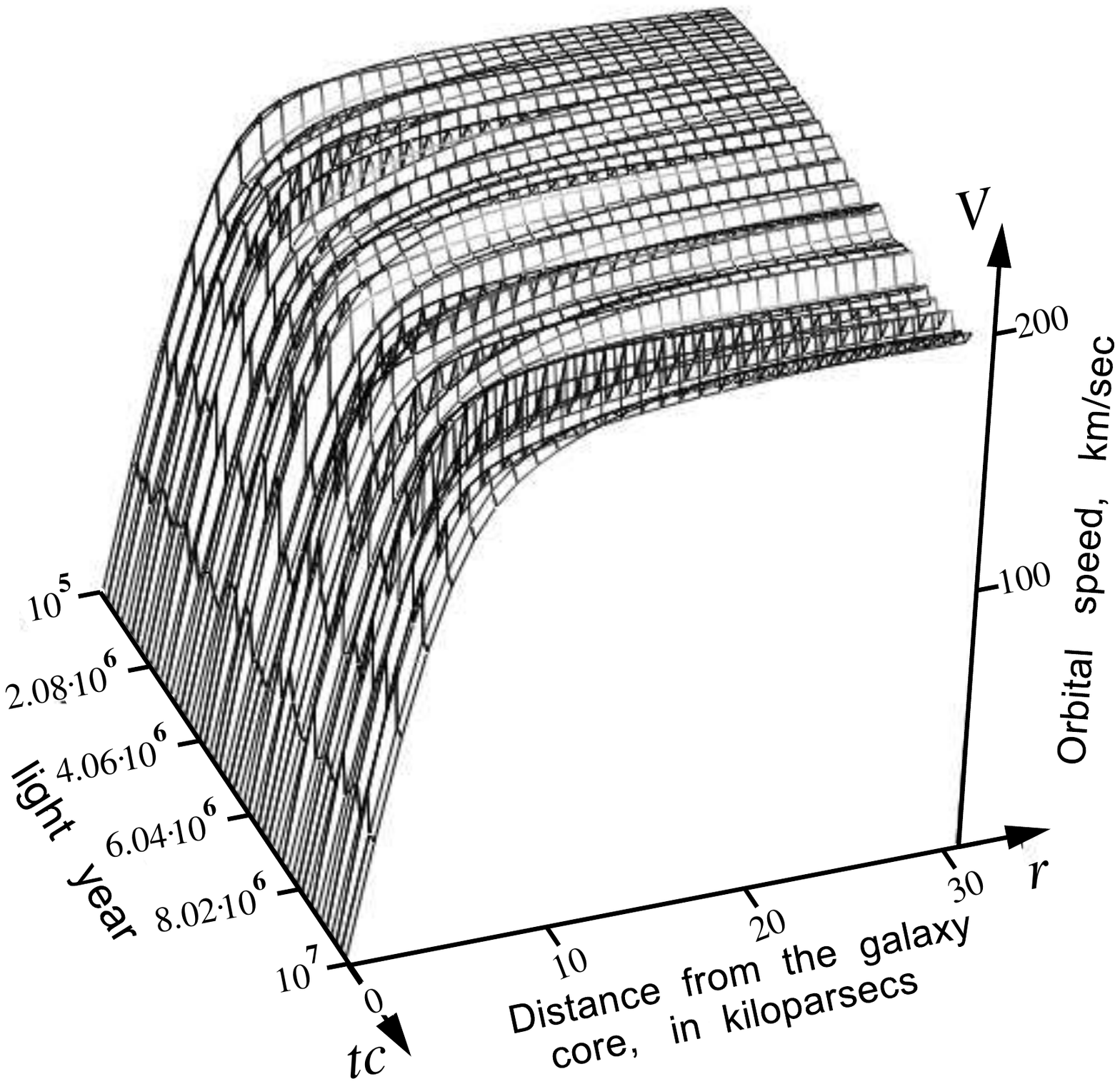}
  \end{picture}
  \caption{
  Orbital speed ${\mathit V}$  is a function of 
 the radius $r$ from the galactic center (in kiloparsec) 
 and time $t$ (in light years).
  }
\label{fig=5}
\end{figure}
To see the formation of the flat profile we need to perform computations of sets collected from modes~(\ref{eq=14}),
 $n = 1, 2, \cdots, N$, at different $N$.
 Let us substitute the expression~(\ref{eq=14}) into the integral~(\ref{eq=9}). After computing it we get
  the following view of the denominator ${\mit\Sigma}(\nu,t,\sigma)$:
\begin{equation}
  {\mit\Sigma}_{n}(t) = {{c^2}\over{\Omega_{n}^{2}}}(\sin(\Omega_{n} t) + \sigma_{0}^{2}).
\label{eq=15}
\end{equation}
 Here the coefficient $\sigma_{0}$ has to be more than 1.
 The fact that $\sigma_{0}$ exceeds~1, prevents possible loss of stability of the vortex (galaxy in our case),
  since $\sin(\Omega_{n}t)+\sigma_{0}^{2}$ is always positive.

Since $\Omega_{n}\sim n^{-1}$, the coefficient ${\mit\Sigma}_{n}\sim c^2/\Omega_{n}^2$ tends to infinity as $\Omega_{n}$ goes to zero with growth of $n$. From here it follows, that the expression $1-\exp\{-r^2/4{\mit\Sigma}_n\}$ reaches~1 the more slowly with increasing $r$, the larger is ${\mit\Sigma}_n$. A set of the coefficients ${\mit\Sigma}_n$ for $n=1,2, \cdots$ can give output to the flat profile.

In statistical mechanics, as in quantum mechanics, one computes the canonical partition function for all energetic states of a system being in equilibrium with its environment. So, we should compute an analogous sum of possible orbital speeds of the galaxy for all entangled modes on which the baryonic matter is allowed to  exchange fluctuations with the physical vacuum and the zero-point fluctuating background.
Our statistical sum in that case becomes
\begin{equation}
 {\mathit V}(r,t) = {{\mit\Gamma}\over{2r}}
 \sum\limits_{n=1}^{N}
 \Biggr(
 1 -\exp\Biggr\{
                      -{{r^2} \over {4{\mit\Sigma}_{n}(\,t\,)}}
             \Biggl\}
 \Biggl).
\label{eq=16}
\end{equation}
   The orbital speed  ${\mathit V}(r,t)$ versus $r$ and $t$  
   is shown in~Fig.~\ref{fig=5}.
  Here for evaluated calculations we used ${\mit\Gamma}= 3\!\cdot\!\!10^{25}$ m$^2$/s
    and the angular frequency $\Omega_{n}$ ranges from $10^{-11}$ s$^{-1}$ to $1.667\!\cdot\!10^{-13}$~s$^{-1}$
     as $n$ runs from 1 to 60.
  The angular frequencies are  extremely small.
   While the wavelengths, $\lambda_{n}=c/\Omega_{n}$,  are in the range  from $0.97$ to $58.3$~kpc.
   These oscillating modes cover areas from the galactic core up to the sizes of the galaxy itself. 

  Fig.~\ref{eq=5}  shows also that the orbital speed experiences small fluctuations in time, resembling the breathing of the galaxy. This breathing is caused by the exchange of vortex energy with the physical vacuum on the ultra-low frequencies $\Omega_{n}$. In other words, the galaxy breathes in response to an energy exchange with the vacuum.
It should be noted here, that the trembling of galaxy with the $1/f$ spectrum most likely emulates music of the heaven spheres (Fractal Memory~\citep{ARtroniks}), than the breathing.
     
 \section{\label{sec=3}Conclusion}  

The flat profile can be achieved by an exchange of the orbital energy of the baryonic matter with the quantum vacuum zero-point energy. This energy exchange is represented through a change of sign in the viscosity of the special superfluid medium, the physical vacuum. In this time-averaged function the viscosity vanishes on average, but its variance is not zero.
  
 The exchange takes place on the ultra-low frequencies. Recalculating the wavelengths of these fluctuations we find that they cover sizes ranging from the galactic core up to the radius of the galaxy and more.
 
 Let us take a look at an extreme point suggested by arrow c in Fig.~\ref{fig=4}. 
 We note that the frequency $\Omega_{\rm c}$ is $2.2\cdot10^{-18}$ 1/s.  
 We find $c/\Omega_{\rm c}\approx 1.36\cdot10^{26}$~m, 
 which is almost close to the Compton wavelength evaluated for the visible universe in~\citep{GoldhaberNieto2010}.
 This corresponds to the radius of the Hubble sphere $r_{_{HS}}=c/H_{0}$,
 which is about $4.4\cdot10^{3}$ Mpc, or about 14 billions  light years 
 (the Hubble constant $H_{0} = 67.8$ km/(s$\cdot$Mpc)). 
This corresponds with the radius of the observable universe.
  
 
We could continue this summation~(\ref{eq=16}) up to the point $\Omega_{\rm c}=2.2\cdot10^{-18}$ 1/s. 
This allows us to affirm that the observable universe rotates about some center with an orbital speed, which has a flat profile through enormous distances. Excepting a central region where the orbital speed grows from zero to the maximal value corresponding to the profile level. This rotation possibly takes place around the richest SuperCluster in the Sloan Great Wall, SCl~126, and especially around its core, resembling a very rich filament~\citep{EinastoEtAl2011}.

On these cosmological scales we can evaluate a mass of a graviton, with a wavelength that is commensurable with the radius of the universe stated above. An extremal mass of the axion-like particle for the observable universe is
\begin{equation}
 m_{g} = {{\hbar\Omega_{\rm c}}\over{c^2}} \approx 2.6\cdot10^{-69}~{\rm kg}.
\label{eq=17}
\end{equation}
This value finds a good agreement with evaluation that comes from the model of 
the holographic screen to be the boundary of the visible Universe ~\citep{MureikaMann2011}.
This evaluation is in agreement with the graviton mass given in~\citep{Tripple2013}.
Ultra-light dark matter particles producing out of vacuum has been predicted in the work of~\citep{ChefranovNovikov2010}.

The frequency spectrum obeys the $1/f$-law (the flicker-noise law). 
As the frequency goes to zero the viscosity coefficient~(\ref{eq=14}) approaches infinity. 
We may imagine that on these limiting frequencies
 the universe can have positive viscosity during one stage of its evolution and negative viscosity during other. When the viscosity is negative, the physical vacuum returns the energy to the baryonic matter. This means, that the baryonic matter is created out of nothing and we observe the expansion of the universe. The most likely candidates for creating baryonic matter may be heavy stars and black holes jets.
On the other hand, when the viscosity is positive baryonic matter loses energy. This means that baryonic matter gradually disappears and the universe shrinks towards zero. Theoretically, only a perfect vacuum remains.
One can observe that such cyclic variations of the baryonic matter and the emtpy space find the  correlations with the Baum-Frampton model~\citep{BaumFrampton2007, BaumFrampton2008, BaumEtAl2008}. According to this model the Universe periodically comes back to "empty space" and returns the baryonic matter.
This research confirms also that a spatially flat Universe exists in time without any
singularities such as the Big Bang.
After some time, about 38 billion years~\citep{ChefranovNovikov2010}, the solution becomes unstable and characterizes the inverse process of absorption of dark particles using the vacuum in the compression mode of the Universe. 
Gurzadyan and Penrose~\citep{GurzadyanPenrose2010} also have described in 2010 a scenario of cyclic repeating of  evolution between a series of big bangs, each of which creates a new the Universe history.

\begin{acknowledgments}

  The author thanks
  Mike Cavedon,
  Thad Roberts, 
  and Denise Puglia
  for  valuable remarks and propositions.

\end{acknowledgments}


\begin{thebibliography}{10}

\bibitem{Whittaker1990}
E.~T. Whittaker.
\newblock {\em History of the Theories of Aether \& Electricity}.
\newblock Dover Publications, 1990.

\bibitem{Fedi2016}
M.~Fedi.
\newblock Gravity as a fluid dynamic phenomenon in a superfluid quantum space.
  {Fluid} quantum gravity and relativity. {HAL Id: hal-01248015}, 20 Apr 2016.

\bibitem{SorliEtAl2016}
A.~Sorli, V.~Koroli, A.~Nistreanu, and D.~Fiscaletti.
\newblock {Cosmology of Einstein�s NOW}.
\newblock {\em American Journal of Modern Physics}, 5(4-1):1--5, 2016.

\bibitem{DwornikEtAl2014a}
M.~Dwornik, Z.~Keresztes, and L.~{\'A} Gergely.
\newblock Rotation curves in {Bose-Einstein Condensate Dark Matter Halos}.
\newblock In A.~Nakajima N.~Kinjo, editor, {\em Recent Development in Dark
  Matter Research}, chapter~6, pages 195--219. Nova Science Publishers, 2014.

\bibitem{DwornikEtAl2014b}
M.~Dwornik, Z.~Keresztes, and L.~{\'A} Gergely.
\newblock {Bose-Einstein Condensate Dark Matter Halos} confronted with galactic
  observations {arXiv}:1406.0388, 2 Jun 2014.

\bibitem{DasBhaduri2015}
S.~Das and R.~K. Bhaduri.
\newblock Dark matter and dark energy from {Bose-Einstein} condensate.
\newblock {\em Class. Quant. Grav.}, 32:105003 (6 pp), 2015.

\bibitem{Ardey2006}
A.~Arbey.
\newblock Dark fluid: A complex scalar field to unify dark energy and dark
  matter.
\newblock {\em Phys. Rev. D}, 74:043516, 2006.

\bibitem{ChefranovNovikov2010}
S.~G. Chefranov and E.~A. Novikov.
\newblock Hydrodynamic vacuum sources of dark matter self-generation in an
  accelerating universe without a {Big Bang}.
\newblock {\em Journal of Experimental and Theoretical Physics},
  111(5):731--743, 2010.

\bibitem{BoehmerHarko2007}
C.~G. Boehmer and T.~Harko.
\newblock Can dark matter be a {Bose-Einstein} condensate?
\newblock {\em JCAP}, 2007(06):025, 2007.

\bibitem{HarkoMocanu2012}
T.~Harko and G.~Mocanu.
\newblock Cosmological evolution of finite temperature {Bose-Einstein
  Condensate} dark matter.
\newblock {\em Phys. Rev.}, D85:084012, 2012.

\bibitem{AlbaretiEtAl2014a}
F.~D. Albareti, J.~A.~R. Cembranos, and A.~L. Maroto.
\newblock Vacuum energy as dark matter.
\newblock {\em Phys. Rev. D}, 90:123509, 2014.

\bibitem{Hajdukovic2011a}
D.~S. Hajdukovic.
\newblock Is dark matter an illusion created by the gravitational polarization
  of the quantum vacuum?
\newblock {\em Astrophys. Space Sci.}, 334:215--218, 2011.

\bibitem{Rubin2004}
V.~C. Rubin.
\newblock A brief history of dark matter.
\newblock In Livio M., editor, {\em The Dark Universe: Matter, Energy and
  Gravity}, Symposium Series: 15, pages 1--13. Cambridge University Press,
  Cambridge, 2004.

\bibitem{deBlokEtAl2001}
W.~J.~G. de~Blok, S.~S. McGaugh, and V.~C. Rubin.
\newblock High-resolution rotation curves of low surface brightness galaxies.
  {II.} {M}ass models.
\newblock {\em The Astronomical Journal}, 122:2396--2427, 2001.

\bibitem{PlanckCo2015}
{Planck Collaboration (260 co-authors)}.
\newblock {Planck 2015 results. XIII. Cosmological parameters}.
  {arXiv:1502.01589}, 6 Feb 2015.

\bibitem{Parkhomov1998}
A.~G. Parkhomov.
\newblock Dark~matter:~its~role~in~cosmo-terrestrial interactions.
\newblock {\em Consciousness~and~physical~reality}, 3(6):24--35, 1998.
\newblock In Russian.

\bibitem{Sbitnev2015c}
V.~I. Sbitnev.
\newblock Hydrodynamics of the physical vacuum: dark matter is an illusion.
\newblock {\em Mod. Phys. Lett. A}, 30(35):1550184 (16 pages), 2015.

\bibitem{Sbitnev2015b}
V.~I. Sbitnev.
\newblock Physical vacuum is a special superfluid medium.
\newblock In {Prof. Pahlavani, M. R.}, editor, {\em Selected Topics in
  Applications of Quantum Mechanics}, chapter~12, pages 345--373. InTech,
  Rijeka, 2015.

\bibitem{Sbitnev2016b}
V.~I. Sbitnev.
\newblock {Hydrodynamics of the physical vacuum: I. Scalar quantum sector}.
\newblock {\em Foundations of Physics}, 46(5):606--619, 2016.

\bibitem{Sbitnev2016c}
V.~I. Sbitnev.
\newblock {Hydrodynamics of the physical vacuum: II. Vorticity dynamics}.
\newblock {\em Foundations of Physics}, pages 1--15, 3 2016.

\bibitem{LicataFiscaletti2014b}
I.~Licata and D.~Fiscaletti.
\newblock {\em Quantum potential: Physics, Geometry, and Algebra}.
\newblock Springer, Cham, Heidelberg, N. Y., Dordrecht, London, 2014.

\bibitem{Fiscaletti2012}
D.~Fiscaletti.
\newblock The geometrodynamic nature of the quantum potential.
\newblock {\em Ukr. J. Phys.}, 57(5):560--573, 2012.

\bibitem{AliDas2015}
A.~F. Ali and S.~Das.
\newblock Cosmology from quantum potential.
\newblock {\em Physics Letters B}, 741:276--279, 4 Febeuary 2015.

\bibitem{Nelson1966}
E.~Nelson.
\newblock Derivation of the {S}chr\"{o}dinger equation from {N}ewtonian
  mechanics.
\newblock {\em Phys. Rev.}, 150:1079--1085, 1966.

\bibitem{ClampittJainm2015a}
J.~Clampitt and B.~Jainm.
\newblock Lensing measurements of the mass distribution in {SDSS} voids.
\newblock {\em Mon. Not. R. Astron. Soc.}, 454(4):3357--3365, 2015.

\bibitem{Poluyan2015R}
P.~V. Poluyan.
\newblock {\em Death of the dark matter: philosophical principles of physical
  knowledge}.
\newblock Krasnoyarsk State University, Krasnoyarsk, 2015.
\newblock (In Russian).

\bibitem{ProvenzaleEtAl2008}
A.~Provenzale, A.~Babiano, A.~Bracco, C.~Pasquero, and J.~B. Weiss.
\newblock Coherent vortices and tracer transport.
\newblock {\em Lect. Notes Phys.}, 744:101--116, 2008.

\bibitem{KleinebergFriedrich2013}
K.~K. Kleineberg and R.~Friedrich.
\newblock Gaussian vortex approximation to the instanton equations of
  two-dimensional turbulence.
\newblock {\em Phys. Rev. E}, 87:033007, 2013.

\bibitem{KevlahanFarge1997}
N.~K.-R. Kevlahan and M.~Farge.
\newblock Vorticity filaments in two-dimensional turbulence: creation,
  stability and effect.
\newblock {\em J. Fluid Mech.}, 346(9):49--76, 1997.

\bibitem{MielczarektAl2010}
J.~Mielczarek, T.~Stachowiak, and M.~Szydlowski.
\newblock Vortex in axion condensate as a dark matter halo.
\newblock {\em Int. J. Mod. Phys.}, D19:1843--1855, 2010.

\bibitem{MureikaMann2011}
J.~R. Mureika and R.~B. Mann.
\newblock Does entropic gravity bound the masses of the photon and graviton?
\newblock {\em Mod. Phys. Lett.}, A26:171--181, 2011.

\bibitem{ARtroniks}
ARtroniks.
\newblock {Fractal Memory}.

\bibitem{GoldhaberNieto2010}
A.~S. Goldhaber and M.~M. Nieto.
\newblock Photon and graviton mass limits.
\newblock {\em Rev. Mod. Phys.}, 82:939--979, 2010.

\bibitem{EinastoEtAl2011}
E.~Einasto, L.~J. Liivamagi, E.~Tempel, E.~Saar, E.~Tago, P.~Einasto,
  I.~Enkvist, J.~Einasto, V.~J. Martinez, P.~Heinamaki, and P.~Nurmi.
\newblock {The Sloan Great Wall. Morphology and galaxy content}.
\newblock {\em The Astrophysical Journal}, 736(1):25pp, 2011.

\bibitem{Tripple2013}
S.~Triple.
\newblock A simplified treatment of gravitational interaction on galactic
  scales.
\newblock {\em Journal of The Korean Astronomical Society}, 46(1):41--47, 2013.

\bibitem{BaumFrampton2007}
L.~Baum and P.~H. Frampton.
\newblock Turnaround in cyclic cosmology.
\newblock {\em Phys. Rev. Lett.}, 98:071301, 2007.

\bibitem{BaumFrampton2008}
L.~Baum and P.~H. Frampton.
\newblock Entropy of contracting universe in cyclic cosmology.
\newblock {\em Mod. Phys. Lett.}, A23:33--36, 2008.

\bibitem{BaumEtAl2008}
L.~Baum, P.~H. Frampton, and S.~Matsuzaki.
\newblock Constraints on deflation from the equation of state of dark energy.
\newblock {\em JCAP}, 0804:032, 2008.

\bibitem{GurzadyanPenrose2010}
V.~G. Gurzadyan and R.~Penrose.
\newblock {Concentric circles in WMAP data may provide evidence of violent
  pre-Big-Bang activity}, 16 Nov 2010.

\end{thebibliography}

\end{document}